%% file: main.tex
\documentclass[twocolumn]{aastex63}
\pdfoutput=1 
\usepackage[T1]{fontenc}
\usepackage{amsmath,amstext}
\usepackage{amssymb}
\usepackage[T1]{tipa}
\usepackage{apjfonts} 
\usepackage[figure,figure*]{hypcap}
\usepackage{microtype}
\usepackage{tablefootnote}
\usepackage{booktabs}  
\usepackage{longtable}
\usepackage{hyperref}
\usepackage{url}
\usepackage{CJKutf8}

\newcommand{\frb}{FRB\,20240209A}

\newcommand{\kkoname}{k'ni\textipa{P}atn k'l$\left._\mathrm{\smile}\right.$stk'masqt}

\begin{document}

\shortauthors{Eftekhari et al.}

\shorttitle{A Massive and Quiescent Elliptical FRB Host Galaxy}

\title{The Massive and Quiescent Elliptical Host Galaxy of the Repeating Fast Radio Burst \frb}
\input{affiliations}

\input{authors}

\begin{abstract}
The discovery and localization of \frb\ by the Canadian Hydrogen Intensity Mapping Fast Radio Burst (CHIME/FRB) experiment marks the first repeating FRB localized with the CHIME/FRB Outriggers and adds to the small sample of repeating FRBs with associated host galaxies. Here we present Keck and Gemini observations of the host that reveal a redshift $z=0.1384\pm0.0004$. We perform stellar population modeling to jointly fit the optical through mid-infrared data of the host and infer a median stellar mass log$(M_*/{\rm M_{\odot}})=11.34\pm0.01$ and a mass-weighted stellar population age $\sim11$Gyr, corresponding to the most massive and oldest FRB host discovered to date. Coupled with a star formation rate $<0.36\,{\rm M_{\odot}\ yr^{-1}}$, the specific star formation rate
$<10^{-11.8}\rm\ yr^{-1}$ classifies the host as quiescent. Through surface brightness profile modeling, we determine an elliptical galaxy morphology, marking the host as the first confirmed elliptical FRB host. The discovery of a quiescent early-type host galaxy within a transient class predominantly characterized by late-type star-forming hosts is reminiscent of short-duration gamma-ray bursts, Type Ia supernovae, and ultraluminous X-ray sources. Based on these shared host demographics, coupled with a large offset as demonstrated in our companion paper, we conclude that preferred progenitors for \frb\ include magnetars formed through merging binary neutron stars/white dwarfs or the accretion-induced collapse of a white dwarf, or a luminous X-ray binary. Together with FRB\,20200120E localized to a globular cluster in M81, our findings provide strong evidence that some fraction of FRBs may arise from a process distinct from the core collapse of massive stars.

\end{abstract}


\section{Introduction}
\label{sec:intro}
Investigations into the host galaxies of transient phenomena and their stellar populations have played a key role in our understanding of their progenitors (e.g., \citealt{vandenBergh1991,Berger2009,Fong2010,Lunnan2014,Nugent2022,Gordon2023}). The association of long gamma-ray bursts (GRBs) with low-metallicity star-forming galaxies solidified their connection to massive stellar progenitors \citep{Bloom2002} that trace the brightest regions of their host galaxies \citep{Fruchter2006}. At the same time, a small fraction of short GRBs and Type Ia supernovae (SNe) localized to early-type galaxies with no appreciable star formation immediately implicated older stellar populations for some subset of events, and hence a wide distribution of delay times for their progenitors \citep{Prochaska2006,Totani2008,Fong2013}. 

\begin{figure*}
\center
\includegraphics[width=0.9\textwidth]{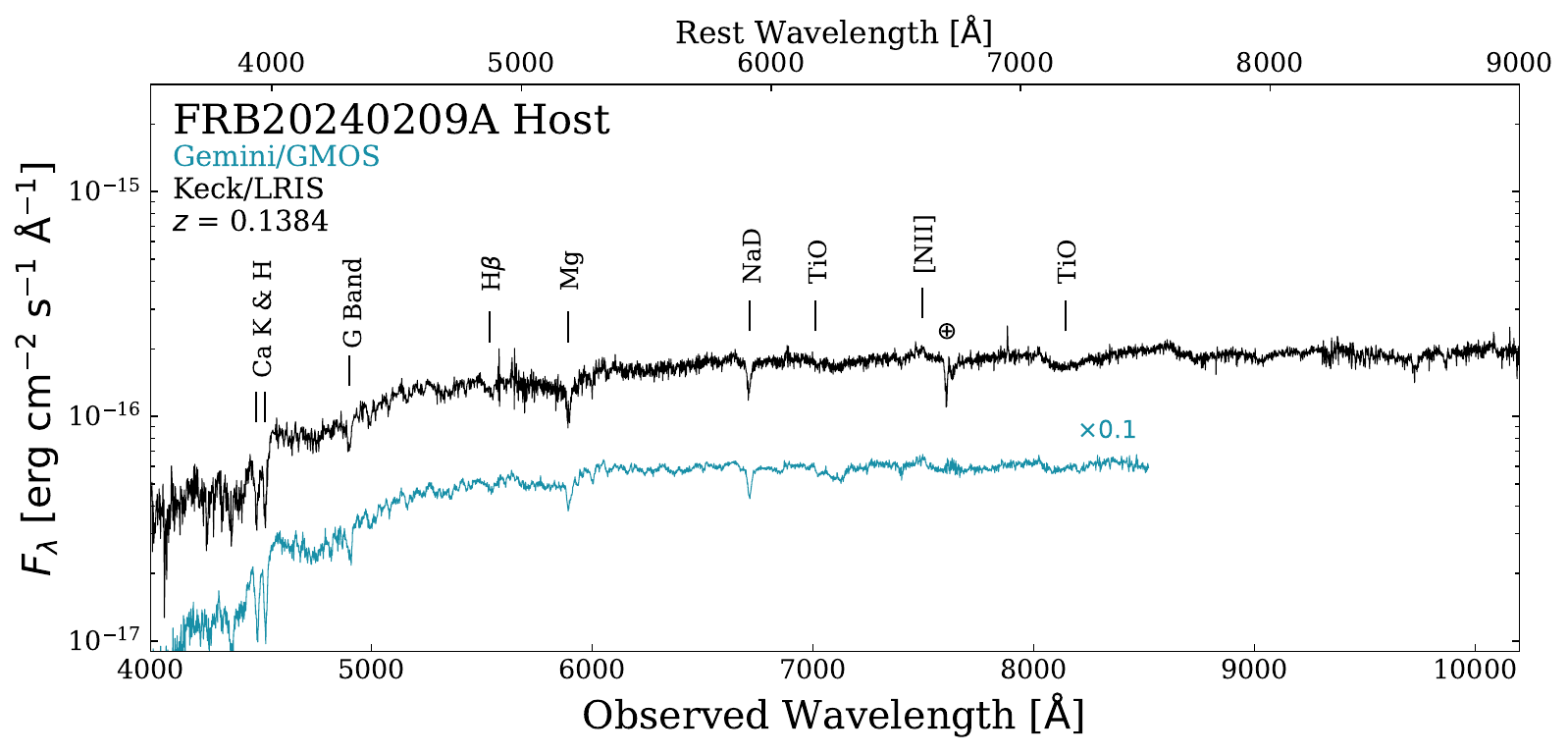}
\caption{Gemini/GMOS and Keck/LRIS spectra of the host galaxy of \frb, where we arbitrarily scale the Gemini/GMOS spectrum on the y-axis for easier visualization. The locations of several observed spectral features that enabled a redshift determination are denoted. Telluric features are marked with $\oplus$.}
\label{fig:spectrum}
\end{figure*}

In contrast, early studies of fast radio burst (FRB) host demographics have demonstrated that the vast majority of events reside in galaxies that are actively star-forming \citep{Heintz2020,Bhandari2022,Eftekhari2023,Gordon2023,Bhardwaj2024,Law2024,Sharma2024} and are often associated on small spatial scales with regions of ongoing star formation \citep{Chittidi2021,Mannings2021,Dong2024}, pointing to a population of young stellar progenitors. FRBs are bright, millisecond duration flares of coherent radio emission originating primarily from  extragalactic distances and whose progenitors otherwise remain unknown \citep{Lorimer2007,Thornton2013}. While the known sample of events exceeds 700 FRBs published to date (e.g., \citealt{Macquart2020,CHIME2020,Law2024,Shannon2024}), the vast majority are poorly localized ($\sim$ arcminutes), precluding robust associations to their host galaxies \citep{Eftekhari2017} except for the most nearby events \citep{Bhardwaj2021,Michilli2023,Ibik2024}. This dearth of FRB host galaxies, coupled with an ostensible dichotomy between $\lesssim 10\%$ of the observed population that exhibit repeat bursts (so-called ``repeaters'') and apparent one-off events \citep{Shannon2018,CHIME2021}, have posed a major barrier towards uncovering their elusive origins and possible progenitor systems.

It is currently widely held that the majority of FRBs are produced by magnetars formed through core-collapse supernovae (CCSNe), in large part due to their preferential occurrence in star-forming environments \citep{Bochenek2021,Gordon2023,Bhardwaj2024,Sharma2024} paired with the discovery of FRB-like events from a known Galactic magnetar \citep{CHIME2020,Bochenek2020}. The observational characteristics of FRBs themselves, including their high brightness temperatures, polarization properties, energetics, burst rates and durations lend further support for a compact object origin \citep{Michilli2018,Dai2021,Pleunis2021,Nimmo2022,Sherman2023,Bera2024,Mckinven2024,Nimmo2024}. On the other hand, a small fraction of events have been localized to quiescent galaxies with typical ages of $\sim 8$ Gyr \citep{Bannister2019,Gordon2023,Sharma2023,Law2024}, indicating that delayed channels are viable for some subset of FRBs \citep{Margalit2019,Pelliciari2023}. Notably, no repeating FRB has been unambiguously associated to a quiescent galaxy thus far. 

In particular, the association of $\lesssim 5\%$ of localized FRBs to massive quiescent galaxies has prompted progenitor theories invoking longer-lived magnetars formed through binary neutron star (BNS) mergers or the accretion-induced collapse (AIC) of a white dwarf (WD) \citep{Margalit2019}. BNS mergers, a small fraction of which may leave behind a stable magnetar remnant (e.g., \citealt{Price2006,Margalit2019b}), should occur in both star-forming and quiescent galaxies, owing to a broad range in delay times between binary formation and eventual merger \citep{Belczynski2006,Zheng2007}. This conclusion is further supported by the observed demographics of short GRB host galaxies \citep{Berger2005,Fong2013,Nugent2022,Jeong2024}, although no definitive evidence for the creation of stable magnetars in this way has been seen \citep{Schroeder2020}. Alternatively, a stable magnetar remnant may form following the collapse of a WD due to accretion from a companion \citep{Brooks2017} or the merger of two WDs \citep{Schwab2016}. Provided that the system is formed in the field, the large average delay time between star formation and AIC/WD merger\textbf{s} would be similarly mirrored in the host demographics. This scenario is in contrast to CCSNe, which predominantly occur in star-forming galaxies due to their massive stellar progenitors.

Perhaps most strikingly, the milliarcsecond localization of the repeating FRB\,20200120E \citep{Bhardwaj2021_m81} to an old ($\sim 9.1$ Gyr) globular cluster (GC) \citep{Kirsten2022} provides conclusive evidence that at least some FRBs formed through delayed channels. Recent work has shown that the centers of GCs that have undergone core-collapse are predominantly composed of white dwarfs and neutron stars especially late in their histories \citep{Ye2019,Kremer2020}, and hence are ideal sites for the dynamical formation of BNS/binary WD mergers or WD AIC \citep{Kremer2021}. Indeed, GCs have been observed to host a wide range of high-energy-emitting sources, ranging from millisecond pulsars \citep{Ransom2008} to low-mass X-ray binaries \citep{Clark1975} and including ultra-luminous X-ray sources (ULXs; \citealt{Maccarone2007,Dage2021}) that have been suggested as potential FRB progenitors \citep{Sridhar2021}. On the other hand, it is difficult to explain the large rate of FRBs in GCs \citep{Rao2024}. In the case of FRB\,20200120E it is furthermore worth noting that deep X-ray limits preclude a ULX origin \citep{Pearlman2023}. Interestingly, roughly one-third of known ULXs are located in early-type elliptical galaxies \citep{Angelini2001,Walton2011,Plotkin2014,Thygesen2023} with low X-ray luminosities that are markedly different from their spiral galaxy counterparts \citep{Irwin2004,Zhang2012}. 

Uncovering a population of FRBs in galaxies without active star formation may therefore implicate one or more of the aforementioned source classes for some fraction of FRB successors. Moreover, with repeating FRBs comprising $\lesssim 20 \%$ of localized sources, it remains unclear whether their environments are statistically distinct from those of non-repeating FRBs (e.g., \citealt{Bhandari2022,Heintz2020,Gordon2023,Bhardwaj2024,Pandhi2024}), and thus whether repeating FRB sources trace unique stellar populations, intensifying the need for larger samples of precisely localized repeating FRBs. The Canadian Hydrogen Intensity Mapping Experiment (CHIME)/FRB project,  which observes the full Northern sky ($\delta > -11^\circ$) daily \citep{CHIME2018}, aims to meet this demand with the addition of the CHIME/FRB Outriggers to the array \citep{Lanman2024}. The first of three outriggers is already operational, enabling very long baseline interferometric (VLBI) localizations of order $\sim 1\arcmin \times 2\arcsec$, sufficient for robust host galaxy identifications at low redshift \citep{CHIME/FRB2024_KK0}.

Here we present the properties of the host galaxy of the repeating \frb, discovered and localized by CHIME/FRB \citep{,Shah2024_atel}. \frb\ was first detected on 2024 February 9. A total of 22 repeat bursts were subsequently discovered, six of which were simultaneously recorded at the CHIME/FRB KKO Outrigger station, enabling an interferometric localization to $\lesssim 2 \arcsec$ and a robust host galaxy association \citep{Shah_R155}. \citet{Shah_R155}  present the discovery and localization of \frb, the properties of the bursts, and a detailed analysis of the host galaxy offset with implications for the FRB progenitor. In this work, we present the spectroscopic redshift and multi-wavelength photometry, examine the overall properties of the host, and contextualize these results within the framework of other transient populations. We note that \citet{Shah_R155} discuss the possibility of an undetected satellite dwarf galaxy within the localization region and conclude that such a scenario would be extreme given existing limits on the luminosity of an undetected host of $L \lesssim 10^7 \ \rm L_\odot$ (a factor of $10 \times$ less luminous than any other FRB host), and the dispersion measure (DM) budget which implies that the host and local environment contribute minimally to the DM, in marked contrast to the sample of FRBs in dwarf galaxies which exhibit substantial host DM contributions \citep{Chatterjee2017,Niu2022,Bhandari2023}. We therefore proceed here under the explicit and well-motivated ($99\%$ association probability; \citealt{Shah_R155}) assumption that the FRB is associated with the bright galaxy located near the localization region. 
 
The structure of the paper is as follows. In Section~\ref{sec:obs}, we discuss our optical follow-up observations as well as the archival data used in our analysis. In Section~\ref{sec:prospector}, we model the host galaxy to derive its stellar population properties. We perform a morphological analysis of the host in Section~\ref{sec:galfit}, and we summarize our results in Section~\ref{sec:conclusions}. Unless otherwise stated, all photometry is reported in the AB magnitude system and corrected for Galactic extinction \citep{Schlafly2011}. Throughout the paper, we use the Planck cosmological parameters for a flat $\Lambda$CDM universe, with $H_{0}$ = 67.66 km s$^{-1}$ Mpc$^{-1}$, $\Omega_m = 0.310$, and $\Omega_{\lambda} = 0.690$ \citep{Planck2020}.

\begin{figure*}
\center
\includegraphics[width=0.9\textwidth]{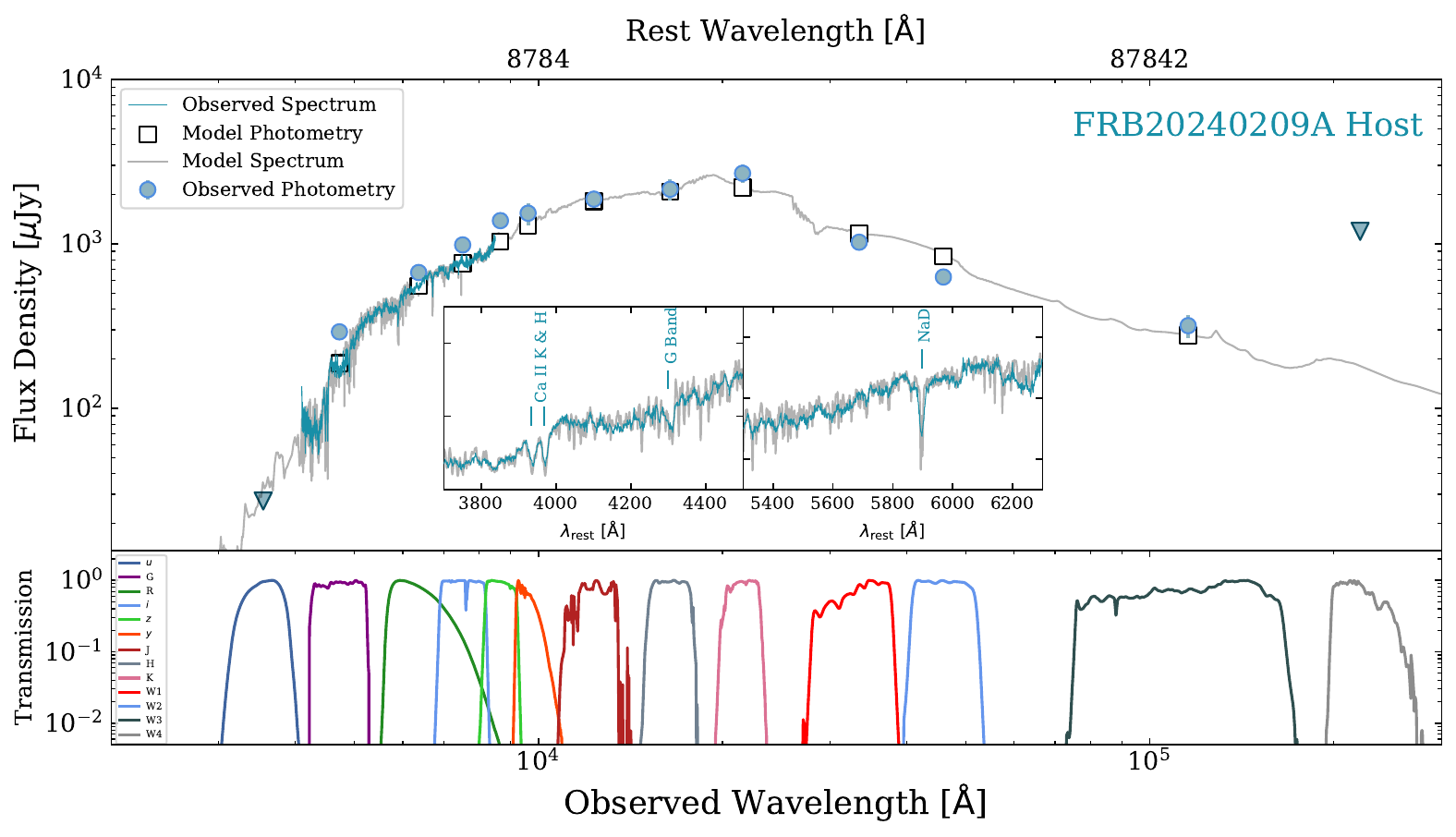}
\caption{Observed SED of the host galaxy of \frb\ including Gemini/GMOS spectroscopy (blue curve) and broadband photometry (blue circles) with the best-fit spectrum (grey curve) and photometry (black squares) from \texttt{Prospector}. The agreement between the observed and modeled spectrum is demonstrated in the zoomed-in inset panels. The bottom panel depicts the transmission filters for each of the photometric points (see also Table~\ref{tab:photom}). Limits correspond to $3\sigma$.}
\label{fig:prospector}
\end{figure*}

\section{Observations}
\label{sec:obs}
\subsection{Photometric Observations}
\label{sec:photometry}

\citet{Shah_R155} present deep $r$-band imaging of the host galaxy and field of FRB\,20240209A\footnote{For host galaxy spectral modeling, we use the Keck R band photometry, taking advantage of the available measurements across multiple Keck filters.}. To further characterize the host, we obtained imaging observations of the field of \frb\ with the 10 m Keck-I Telescope (PI: A.~Miller) in both $G$- and $R$- band on 2024 June 29. The images were processed using the python pipeline \texttt{POTPyRI}\footnote{\url{https://github.com/CIERA-Transients/POTPyRI}}. We performed aperture photometry with a custom script\footnote{\url{https://github.com/charliekilpatrick/photometry}} that utilizes the \texttt{aperture\_photometry} module within the \texttt{photutils} python package \citep{photutils} using a 14\arcsec\, radius aperture ($\sim 5\times r_{\rm eff}$; see Section~\ref{sec:galfit}) centered on the host galaxy. To estimate the background, we created a $5\arcsec$ radius region near the host galaxy and free from any obvious sources. We calibrated all photometry using point sources in common with the PanSTARRS1 (PS1) catalog, and list the final extinction-corrected magnitudes for the host galaxy in Table~\ref{tab:photom}.

\begin{deluxetable}{lccc}
\tablecolumns{4}
\caption{\frb\ Host Galaxy Photometry}
\tablehead{
\colhead{Facility} & 
\colhead{Instrument} &
\colhead{Filter} &
\colhead{Magnitude} \\ 
\colhead{} &
\colhead{} &
\colhead{} &
\colhead{(AB)} 
}  
\startdata 
LCO & Sinistro & $u$ & $>20.31$\\
Keck & LRIS & G & 17.74 $\pm$ 0.08\\
Keck & LRIS & R & 16.84 $\pm$ 0.02\\
Gemini$^{a}$ & GMOS & $r$ & 16.79 $\pm$ 0.02\\
Pan-STARRS & GPC1 & $i$ & 16.42 $\pm$ 0.04\\
Pan-STARRS & GPC1 & $z$ & 16.05 $\pm$ 0.07\\
Pan-STARRS & GPC1 & $y$ & 15.93 $\pm$ 0.17 \\
2MASS &  & $J$ & 15.72 $\pm$ 0.13\\
2MASS &  & $H$ & 15.57 $\pm$ 0.16\\
2MASS &  & $K$ & 15.33 $\pm$ 0.14\\
WISE &  & W1 & 16.38 $\pm$ 0.01\\
WISE &  & W2 & 16.90 $\pm$ 0.02\\
WISE &  & W3 & 16.65 $\pm$ 0.18\\
WISE &  & W4 & $>16.21$ \\
\enddata
\tablecomments{All photometry is corrected for Galactic extinction \citep{Schlafly2011}. Photometric points from 2MASS and WISE are given under catalog designations 2MASX J19192430+8603395 and 
WISEA J191924.09+860339.3, respectively. Limits correspond to $3\sigma$.\\
$^{a}$ Originally presented in \citet{Shah_R155}.}
\label{tab:photom}
\end{deluxetable}

To further assess the level of star formation in the host galaxy, we also observed the field of \frb\ in $u$-band with the Sinistro imagers on the Las Cumbres Observatory (LCO) 1-m telescope network \citep{Brown2013} on 2024 Aug 21 (PI: C.~Kilpatrick). We utilized the processed image from the LCO \texttt{BANZAI} pipeline \citep{McCully2018}, which performs pixel-level corrections and astrometric calibration. We performed flux calibration and photometry of the host galaxy using the same methods as described above. We did not detect the putative host in our $u$-band imaging down to a $3\sigma$ magnitude limit of $>20.31$ mag (extinction-corrected), which we derive using the standard deviation in a source-free annulus centered on the host coordinates. The $u$-band non-detection is consistent with the lack of H$\alpha$ emission in the observed spectrum (see Section~\ref{sec:spec}).

\subsection{Spectroscopic Observations}
\label{sec:spec}

We obtained long-slit optical spectroscopy\footnote{All spectroscopic data were collected by the Fast and Fortunate for FRB Follow-up (F4) collaboration (\url{https://www.frb-f4.org/}) using the \texttt{FFFF-PZ} observation management tool, built and modeled after \texttt{YSE-PZ} \citep{Coulter2022,Coulter2023}.} of the host galaxy of \frb\ with the Low Resolution Imaging Spectrometer (LRIS) mounted on the Keck I telescope at Mauna Kea, Hawaii on 2024 June 29 ( PI: A.~Miller; \citealt{Dong2024_atel}). We took $2 \times180$~s exposures with a 1\arcsec\ slit width using the B400/3400 grism and the R400/8500 grating. We used the Python Spectroscopic Data
Reduction Pipeline (\texttt{PypeIt}, v1.16;  \citealt{Prochaska2020}) to reduce and coadd the spectra using standard reduction techniques. We applied absolute flux callibration using spectrophotometric standard spectra and a telluric correction utilizing the standard atmospheric model grids available in \texttt{PypeIt}. 

To determine the host galaxy redshift and uncertainty, we cross-correlated the spectrum with a 5-Gyr galaxy template for the spectral continuum, described by \citet{BruzualCharlot2003}. We identify prominent spectral features (Figure~\ref{fig:spectrum}), including Ca II H \& K, Mg, NaD, TiO, and a strong 4000\,\AA\, spectral break, corresponding to a redshift $z = 0.1384 \pm 0.0004$ (first reported by \citealt{Dong2024_atel}). We note that this is consistent with the upper limit on the redshift of $z_{\rm max} = 0.19$ as inferred from the FRB's DM \citep{Shah_R155}.

To obtain a high signal-to-noise spectrum for modeling the spectral energy distribution (SED), we also observed the host of \frb\ with the Gemini North Multi-Object Spectrograph (GMOS) in two separate epochs on 2024 July 14 and 2024 Aug 7/8 as part of an ongoing Gemini Large and Long Program (PI: T.~Eftekhari) to conduct host galaxy follow-up of CHIME/FRB Outrigger-localized events. Our first epoch consisted of four 600-s observations with a 1\arcsec\ slit width using the B480 grating and the GG455 blocking filter at central wavelengths of 640 and 650 nm. Our second epoch (eight 900-s exposures) was taken with the same grating and no blocking filter at central wavelengths of 540 and 550 nm. The data were reduced following the same procedure described above for the Keck/LRIS spectroscopy. To obtain a final, flux calibrated spectrum, we coadded the 1D spectra from each individual epoch (see Figure~\ref{fig:spectrum}).

\subsection{Archival Data}

To construct a broadband SED for the host galaxy of \frb\ and facilitate the spectral modeling in Section~\ref{sec:prospector}, we supplement the photometric and spectroscopic observations obtained in this work with archival data from a number of surveys. In particular, we include $izy$ photometry from the Pan-STARRS1 (PS1) $3\pi$ survey \citep{PS1}, $JHK$ photometry from the Two Micron All
Sky Survey (2MASS; \citealt{2MASS}) for catalogued source 2MASX J19192430+8603395, and W1, W2, W3, and W4 photometry from the Wide-field
Infrared Survey Explorer (WISE) All-Sky source catalog \citep{WISE} for the source designated WISEA J191924.09+860339.3. Given the large angular extent of the host galaxy, we utilize the 2MASS Extended Source Catalog \citep{Jarrett2000} for the 2MASS data. For PS1, we perform manual aperture photometry on each of the $izy$ images following the procedure described in Section~\ref{sec:photometry}. Finally, we correct all photometry for Galactic extinction \citep{Schlafly2011} using the \citet{Cardelli1989} extinction law. The final extinction-corrected photometry for each telescope and filter is listed in Table~\ref{tab:photom}.

\section{Host Galaxy Spectral Modeling}
\label{sec:prospector}

\begin{deluxetable}{lcc}
\caption{Derived Host Galaxy Properties of \frb}
\label{tab:pros}
\tablehead{
\colhead{Property} & 
\colhead{Value} &
\colhead{Units}} 
\startdata 
$z$ & $0.1384\pm0.0004$ \\
R.A. (J2000) & 289.85036 $\pm\, 0.00003$  & deg\\
Decl. (J2000) & 86.06090 $\pm\, 0.00003$& deg\\ 
$t_m$ & $11.33^{+0.04}_{-2.92}$ &  Gyr\\
log(M$_*$/M$_{\odot}$) & $11.34 \pm\, 0.01$ &  \\
log(Z$_*$/Z$_{\odot}$) &  $0.19 \pm\, 0.01$ & \\
log(Z$_{\rm gas}$/Z$_{\odot}$) &  $-0.24 \pm\, 0.07$ & \\
A$_V$ & $0.007^{+0.007}_{-0.005}$ & AB mag\\
SFR$_{0--100~Myr}$ &  $<0.36$ & M$_{\odot}$ yr$^{-1}$\\
log(sSFR$_{0--100~Myr})$ & $<-11.8$ & yr$^{-1}$\\
$r_{\rm eff}$ & 7.78 $\pm\, 0.03$ &  kpc\\
$n$ & $4.0\pm0.2$& \\
\enddata
\end{deluxetable}

\subsection{The Old and Quiescent Host of \frb}

We derive the host stellar population properties using the stellar population Bayesian inference code \texttt{Prospector} (\citealt{Johnson2017,Johnson2021}; see Appendix~\ref{sec:appendix_prospector}). We show the observed broadband SED compared to the best-fit \texttt{Prospector} model in Figure~\ref{fig:prospector} and list the inferred best-fit parameters in Table~\ref{tab:pros}. We find a low current star formation rate (SFR; averaged over the last 100 Myr) of SFR$_{\rm{0-100 Myr}} < 0.36 \ \rm M_{\odot} \ yr^{-1}$ and a current stellar mass log$(M_*/M_{\odot}) = 11.36$, corresponding to the most massive FRB host discovered to date. We also infer a mass-weighted stellar population age $t_m \approx 11.33$ Gyr, significantly older than the existing sample of FRB hosts which has $\langle t_m \rangle \approx 5$ Gyr \citep{Gordon2023,Bhardwaj2024,Ibik2024,Law2024,Sharma2024}.

\begin{figure}
\includegraphics[width=\columnwidth]{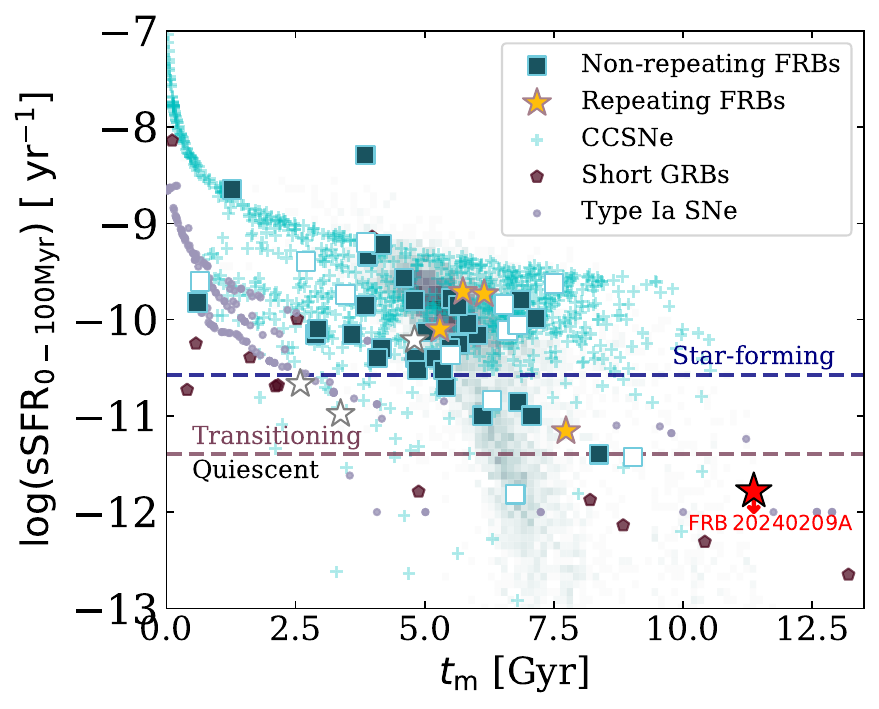}
\caption{Specific star formation rate as a function of mass-weighted age for the host of \frb\ and for a sample of FRB hosts at $z<0.3$ from the literature. We denote FRBs with sSFRs and mass-weighted ages derived in a similar manner as for \frb\ from \citet{Gordon2023} and \citet{Sharma2024} as closed symbols. Also shown are the values for the hosts of CCSNe \citep{Schulze2021}, Type Ia SNe \citep{Neill2009}, and short GRBs \citep{Blanchard2017,Nugent2022}. Open symbols for FRBs and values for SNe are not necessarily derived using the same formalism. Ages for the Type Ia SNe sample are weighted by the host galaxy luminosity and hence represent an underestimate of the mass-weighted age. For comparison, we include the distribution of field galaxies from the COSMOS sample at $z<0.3$ in grayscale.}
\label{fig:sSFR}
\end{figure}

To establish the quiescent nature of the host of \frb, we plot in Figure~\ref{fig:sSFR} the specific star formation rate (sSFR) as a function of mass-weighted age for \frb. We determine \frb's location in this phase-space following the prescription of \citet{Gordon2023} and draw $N=1000$ posterior samples of $\rm SFR_{\rm 0-100\ Myr}$, $M_*$, and $t_m$, ensuring that each of the samples is pulled from the same model posteriors. We next calculate the mass-doubling number $\mathcal{D}(z) = {\rm sSFR} \times t_H(z)$ where $t_H(z)$ is the age of the universe at a redshift $z$ \citep{Tacchella2022}. A galaxy is classified as star-forming, transitioning, or quiescent if $\mathcal{D}(z) > 1/3$, $1/20 < \mathcal{D}(z) < 1/3$, or $\mathcal{D}(z) < 1/20$, respectively. Calculating $\mathcal{D}(z)$ for each of the 1000 model draws, we take the mode to determine \frb's galaxy classification and find that $99\%$ of $\mathcal{D}(z)$ values correspond to the quiescent region of this phase space. The median log(sSFR) from our posterior samples is $-11.8 \ \rm yr^{-1}$ which we adopt as an upper limit given the lack of evidence for star formation in the host spectrum. This marks the host of \frb\ as the first repeating FRB in a quiescent host galaxy.

\begin{figure}
\includegraphics[width=\columnwidth]{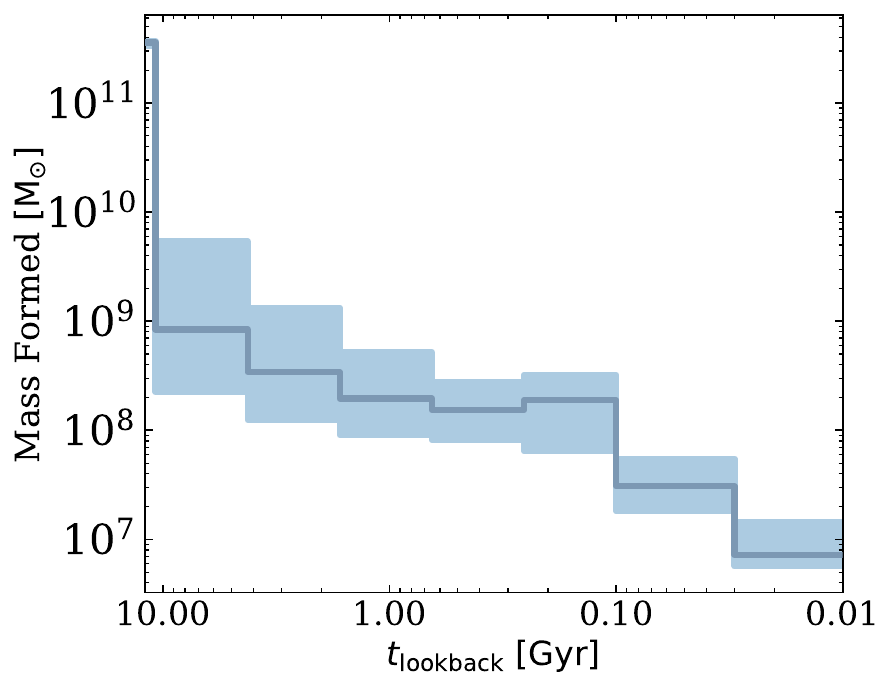}
\caption{Mass assembly history as a function of lookback time for the host galaxy of \frb\ as inferred from the star formation history, where $t_{\rm lookback} = 0$ corresponds to the redshift of the host.}
\label{fig:massformed}
\end{figure}

Finally, in Figure~\ref{fig:massformed}, we plot the stellar mass formed as a function of lookback time for the host of \frb\ as derived from {\tt Prospector}. We find that the bulk of the mass was formed in the first $\sim$~Gyr (by $t_{\rm lookback} \sim 10$ Gyr), indicative of star-formation quenching early in the galaxy's lifetime. Indeed, most massive elliptical galaxies undergo a suppression of star-formation at early times due to AGN feedback \citep{Springel2005}. The surface brightness distribution that we will derive in Section~\ref{sec:galfit} for the host, which indicates a core-S\'ersic profile as expected for quenched systems \citep{Kormendy2009}, and the detection of radio emission coincident with the host galaxy center \citep{Law2024} possibly indicative of an AGN origin, further support this picture.

\subsection{A Comparison to Other FRBs and Transient Classes}

From Figure~\ref{fig:sSFR}, it is evident that the host of \frb\ lies in a unique region of the sSFR-age phase-space relative to other FRBs at $z<0.3$ \citep{Bhardwaj2021,Gordon2023,Bhardwaj2024,Ibik2024,Law2024,Sharma2024}. Several non-repeating FRBs are also quiescent based on this classification scheme, including FRBs 20210807D, 20220509G, and 20221012A \citep{Gordon2023,Sharma2023,Law2024}. While FRB\,20220509G was initially classified as an early-type elliptical galaxy based on publicly available PS1 images \citep{Sharma2023,Connor2023}, deep optical imaging later revealed a disk morphology \citep{Bhardwaj2024}. Similarly, for FRB\,20210807D, we note the clear presence of spiral arms and the proximity of the FRB to the central region of the galaxy \citep{Shannon2024}. For the repeating FRB\,20180916B, the milliarcsecond localization \citep{Marcote2020} is near a star-forming region along clear spiral arms within its host \citep{Tendulkar21}. Thus, although the host is a transitioning galaxy, it stands in stark contrast to \frb's large offset from its host galaxy center and lack of coincident sub-structure in deep imaging \citep{Shah_R155}. The remaining sample of repeating FRB hosts at $z<0.3$ are concentrated around the star-forming locus in Figure~\ref{fig:sSFR}.

\begin{figure}
\includegraphics[width=\columnwidth]{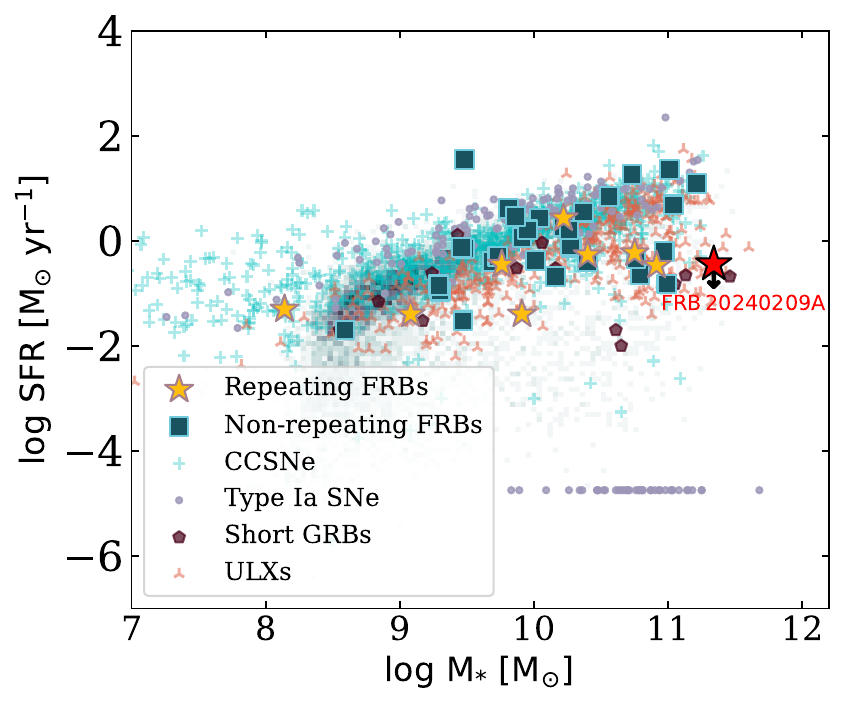}
\caption{Star-formation rates and stellar masses for FRB host galaxies, including \frb, at $z<0.3$ plotted against field galaxies from the COSMOS sample. For comparison, we also include the host galaxies of CCSNe \citep{Schulze2021}, Type Ia SNe \citep{Lampeitl2010}, short GRBs \citep{Blanchard2017,Nugent2022,Jeong2024}, and ULXs \citep{Kovlakas2020}. Type Ia hosts classified as "passive" galaxies in \citet{Lampeitl2010} owing to a lack of recent star formation activity are arbitrarily plotted at log($\rm SFR/M_{\odot} \ yr^{-1})= -4.75$. All other values plotted represent solid SFR measurements.}
\label{fig:sfms}
\end{figure}

\begin{figure*}
\includegraphics[width=\textwidth]{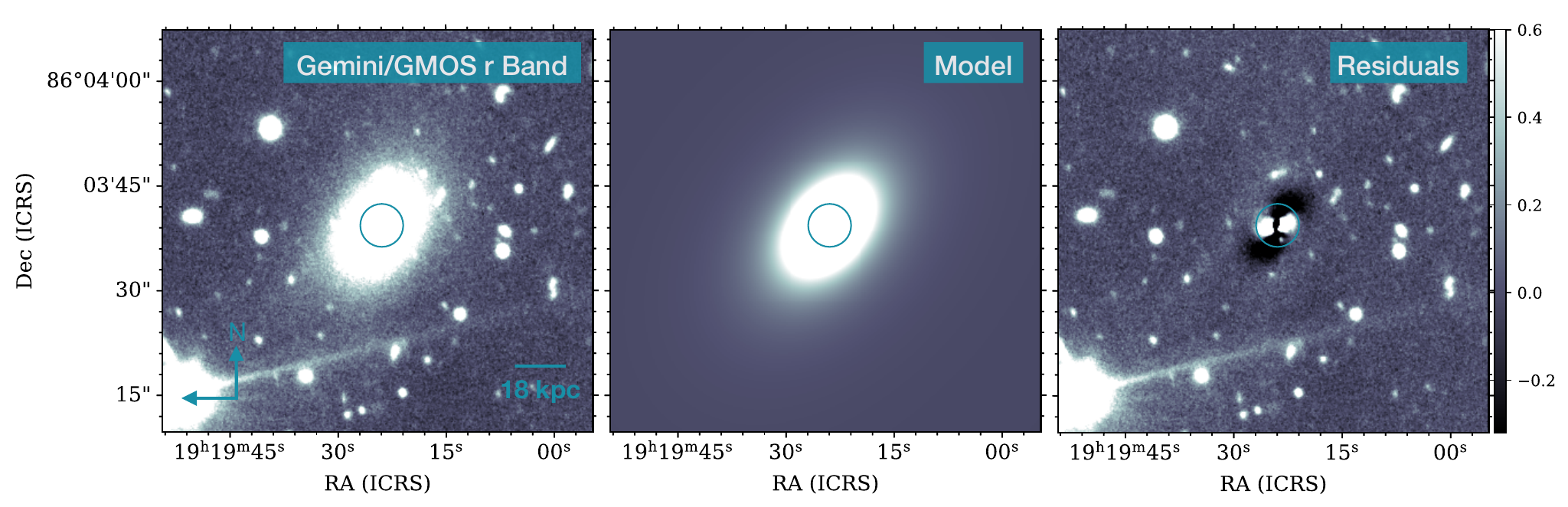}
\caption{Results of our \texttt{GALFIT} modeling of the host of \frb, where we subtract a single-component S\'ersic profile (middle panel) from a Gemini/GMOS $r$-band image (left panel; originally presented by \citealt{Shah_R155}) to derive a residual image (right panel). The circular region with radius $r_e$ illustrates the best-fit effective radius $r_e = 7.78 \pm 0.03$ kpc. Each image is oriented with north up and east to the left.}
\label{fig:galfit}
\end{figure*}

In Figure~\ref{fig:sfms}, we plot the star formation rate and stellar mass of the host of \frb\ against the values for other FRB hosts at $z<0.3$ from the literature. For the host of FRB\,20200120E (M81), we report the H$\alpha$ SFR and galaxy stellar mass from \citet{Gordon2004} and \citet{deBlok2008}, respectively, of $\rm log(SFR) = -0.47 \ \rm M_{\odot} \ yr^{-1}$ and $\rm log(M_{*}) = 10.91 \ M_{\odot}$. While the bulk of FRB hosts lie along the star-forming main sequence (SFMS) \citep{Gordon2023}, a small subset, including \frb, are offset below the SFMS and located in the region corresponding to quiescent galaxies, consistent with its quiescent classification above. A single repeating FRB (FRB\,20200120E in the nearby M81 galaxy) is similar to \frb\ in terms of its offset away from the SFMS and location near the quiescent cloud \citep{Bhardwaj2021_m81}. Indeed, while M81 continues to form stars in its spiral arms at a modest rate, its central bulge is dominated by a very old stellar population. Unlike M81 which exhibits prominent spiral arms, however, the host of \frb\ does not show evidence for on-going star formation anywhere within the galaxy based on the optical spectrum (Section~\ref{sec:obs}).

To compare the the host of \frb\ to classes of known transients, we include in Figure~\ref{fig:sSFR} the sSFRs and ages for the host galaxies of CCSNe \citep{Schulze2021}, Type Ia SNe \citep{Lampeitl2010}, and short GRBs \citep{Blanchard2017,Nugent2022}. The CCSNe sample is derived from the Palomar Transient Factory survey and includes a total of 857 hosts spanning 12 supernova sub-classes, including both hydrogen-rich and hydrogen-poor events as well as superluminous supernovae. For Type Ia SNe, we include 162 hosts at $z<0.21$ from the SDSS-II Supernova Survey \citep{Neill2009}\footnote{Measurements for the Type Ia host ages are light-weighted (versus mass-weighted), and thus skew younger due to the greater emphasis on stellar light from young stars \citep{Conroy2013}.}. We also include 11 short GRB hosts at $z<0.3$ from the Broadband Repository for Investigating Gamma-ray burst Host Traits
(BRIGHT)\footnote{\url{https://bright.ciera.northwestern.edu/}} catalog \citep{Nugent2022} and supplement this sample with the elliptical host galaxy of the binary neutron star merger GW170817 (NGC 4993) with log($\rm SFR/M_{\odot} \ yr^{-1})= -2$ and log($\rm M_{*}/M_{\odot}) = 10.9$ \citep{Blanchard2017}. Finally, as a background sample, we include the distribution of field galaxies at $z<0.3$ from the COSMOS catalog \citep{Laigle2016} modeled with \texttt{Prospector} \citep{Leja2020} using a prescription similar to the one described here.

We find that the \frb\ host is comparable to those of Type Ia SNe and short GRBs in terms of its sSFR and stellar population age, although the hosts of both transient classes also extend into the star-forming region. The host of \frb\ does not resemble the large majority of the CCSNe host population. Indeed, 2\% of CCSNe exhibit sSFRs $\lesssim 10^{-11} \ \rm yr^{-1}$ \citep{Schulze2021} and only $0.9\%$ at $z<0.3$ have both sSFRs and ages within $\sim 0.5$~dex of \frb. We note that this small fraction of CCSNe in quiescent galaxies may implicate binary systems involving massive stars which explode as CCSNe only after merging, and hence long delay times relative to star formation \citep{Zapartas2017}. 

In terms of its SFR and stellar mass, \frb\ is most similar to the hosts of short GRBs\footnote{Here we supplement the short GRB sample described earlier with four hosts at $z<0.3$ from \citet{Jeong2024}.} which exhibit a larger quiescent fraction relative to CCSNe. Indeed, roughly $10\%$ of short GRB hosts are quiescent \citep{Nugent2022}, broadly consistent with theoretical expectations \citep{Chu2022,Santoliquido2022}. We similarly find a small fraction of ULX hosts (compiled using the \textit{Chandra Source Catalog 2.0}; \citealt{Kovlakas2020}) with SFRs and stellar masses comparable to the host of \frb. While the bulk of ULXs are typically found in late-type galaxies \citep{Roberts2000} and are often associated with star-forming regions within their hosts \citep{Gao2003}, a non-negligible fraction reside in early-type galaxies \citep{Angelini2001}. We note that a subset of Type Ia SNe also occurs in massive quiescent galaxies with little to no ongoing star-formation \citep{Lampeitl2010,Smith2012}. These galaxies are represented in Figure~\ref{fig:sfms} with arbitrary values of log($\rm SFR/M_{\odot} \ yr^{-1})= -4.75$ following \citet{Lampeitl2010}.

\section{Host Galaxy Morphological Analysis}
\label{sec:galfit}
\subsection{Evidence for an Elliptical Host Galaxy}

To characterize the morphology of \frb's host galaxy, we perform a S\'ersic (\citeyear{Sersic1963}) profile fit using the \texttt{GALFIT} software package \citep{Peng2002} and the Gemini/GMOS $r$-band image from \citet{Shah_R155} for the analysis (see Appendix~\ref{sec:appendix_galfit}). The Gemini/GMOS image, S\'ersic model, and residual image are shown in Figure~\ref{fig:galfit}. We find that the host galaxy is well-characterized by a S\'ersic index $n = 4.0 \pm 0.2$ and an effective half-light radius $r_e = 7.78 \pm 0.03 \rm \ kpc$. The $n=4$ S\'ersic index is consistent with a de Vaucouleurs (\citeyear{deVaucouleurs1959}) surface brightness profile, characteristic of elliptical galaxies. We note that the residual structure evident in Figure~\ref{fig:galfit} is not uncommon for bright elliptical galaxies, where the inner galaxy profile is best described by an additional power-law component, requiring a modification of the standard S\'ersic profile \citep{Graham2003,Trujillo2004}. The presence of such a component is generally attributed to a deficit of stellar mass in the galaxy center, or depleted core, driven by the merger of two supermassive black holes and their ensuing three-body encounters \citep{Milosavljevic2001,Merritt2006}. This scenario is furthermore consistent with the discovery of bright radio emission coincident with the host galaxy center \citep{Law2024}, as such core-S\'ersic galaxies are more likely to host radio-loud AGN \citep{Capetti2005,Richings2011}. Finally, for comparison, we perform isophotal fitting using the \texttt{isophote.Ellipse} function within the \texttt{PHOTUTILS} package \citep{photutils}. We do not include any sinusoidal components in our model that can account for spiral features. Our fit reveals a clean subtraction and no significant residual structure, consistent with an elliptical morphology and no detected spiral arms.

\begin{figure}
\includegraphics[width=\columnwidth]{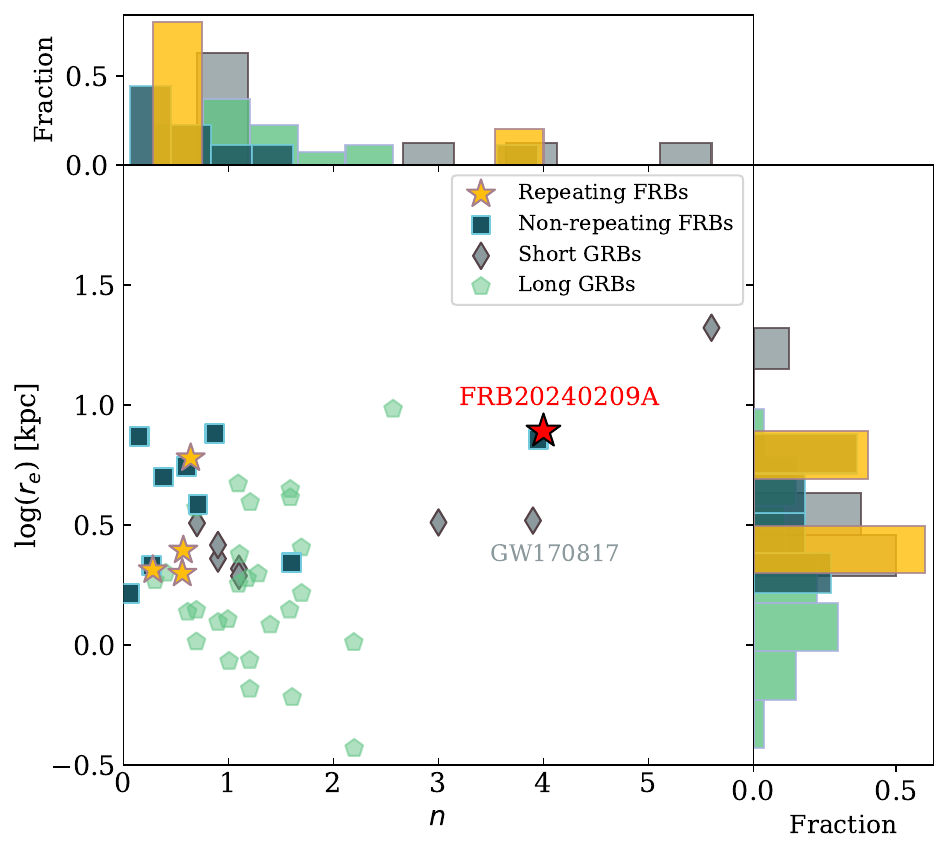}
\caption{Effective radius as a function of S\'ersic index for a sample of FRB hosts from the literature \citep{Heintz2020,Mannings2021,Bhandari2022,Dong2024} compared to the values that we derive for \frb\ in Section~\ref{sec:galfit}. Also shown for comparison are the values derived from HST observations of long GRBs \citep{Wainwright2007} and short GRBs \citep{Fong2010}, where we supplement the short GRB sample with GW170817 \citep{Blanchard2017}.
}
\label{fig:reff}
\end{figure}

In Figure~\ref{fig:reff}, we plot the effective radius and S\'ersic index for \frb\ along with those for a sample of FRB hosts from the literature \citep{Heintz2020,Mannings2021,Bhandari2022,Dong2024,Woodland2024}. We additionally include for comparison the distributions for the host galaxies of both short- \citep{Fong2010} and long-duration \citep{Wainwright2007} GRBs as derived from HST observations. We find that for FRB hosts where similar morphological analyses have been applied, the vast majority of hosts exhibit S\'ersic indices $n < 1.5$, corresponding to exponential disk profiles. \frb\ is a clear outlier in this regard, and is comparable only to the host galaxy of the non-repeating FRB\,20181112A that exhibits a S\'ersic index of $n\sim 4$ \citep{Heintz2020}. Unlike the host of \frb, however, the host of this non-repeating FRB shows clear evidence for on-going star formation \citep{Prochaska2019}, in stark contrast to the quiescent host of \frb. Compared to the hosts of long and short-duration GRBs, \frb\ most closely resembles a subset of short GRBs with de Vaucouleurs profiles ($n > 3$), while the hosts of long GRBs are concentrated around a median value $\langle n \rangle \sim 1.2$ and hence are classified as exponential disks. 

In terms of its effective radius, the host of \frb\ is among the largest FRB hosts with $r_e = 7.78 \pm 0.03 \rm \ kpc$. The hosts of repeating FRBs are on average much smaller, with the exception of FRB\,20180916B in a nearby spiral galaxy with $r_e = 6$ kpc \citep{Marcote2020}. Notably, the hosts of long GRBs extend to much smaller sizes than most FRB hosts \citep{Fong2010}, suggesting that FRBs are unlikely to originate from the same types of hosts as long GRBs. We therefore find that the host of \frb\ most closely resembles a subset of short GRB hosts in terms of its size and morphology.

\subsection{WISE mid-IR Color Diagnostics}
\label{sec:wise}

As an additional qualitative metric to assess the morphology of \frb's host, we examine its location on a WISE mid-IR color-color diagram \citep{Wright2010}. The $W2 - W3$ colors in particular delineate between spheroidal/elliptical galaxies with little on-going star formation ($W2-W3<2$) and spiral galaxies/star-forming disks where an elevated abundance of polycyclic aromatic hydrocarbons (PAHs) results in redder colors \citep{Jarrett2017}. In the vertical direction, mid-IR colors $W1 - W2 >0.8$ indicate the presence of an AGN-heated dusty torus \citep{Stern2012}.

\begin{figure}
\includegraphics[width=\columnwidth]{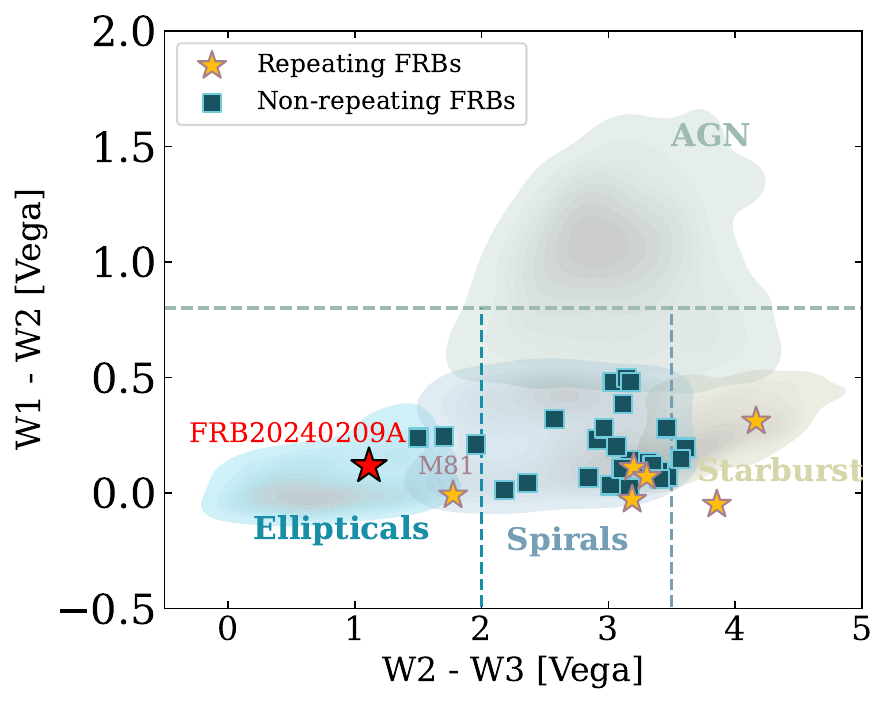}
\caption{\textit{WISE} mid-infrared colors for the host of \frb\ compared to the hosts of repeating (yellow stars) and non-repeating (teal squares) FRBs with \textit{WISE} detections. Shown for comparison are the KDE distributions for the galaxy sub-samples from the MIXR catalog \citep{Mingo2016}. The horizontal dashed line at $W1-W2 = 0.8$ mag denotes the minimum threshold for AGN-dominated galaxies \citep{Stern2012}, while vertical lines mark the division between spheroidal/early-type galaxies, intermediate disks/spiral galaxies, and star-forming/late-type disks \citep{Jarrett2017}.}
\label{fig:wise}
\end{figure}

In Figure~\ref{fig:wise}, we plot the host of \frb\ alongside the sample of FRB hosts with detections in ALLWISE \citep{Gordon2023,Bhardwaj2024,Ibik2024,Law2024,Sharma2024} on a mid-IR color-color diagram. For comparison, we also include the galaxy subsamples from the MIXR catalog \citep{Mingo2016}, comprising detected sources in the mid-IR (ALLWISE), X-rays (3XMM DR5) and radio (FIRST/NVSS). The MIXR sample is broadly divided into four galaxy subclasses (AGN, starburst, spirals, and ellipticals) based on WISE color selections. 

We find that the host of \frb\ occupies a distinct region of this phase-space relative to the vast majority of FRB hosts and corresponding to elliptical galaxies. This is consistent with the surface brightness profile inferred from our \texttt{GALFIT} modeling in Section~\ref{sec:galfit}. With the exception of FRB\,20200120E localized to the nearby M81 galaxy \citep{Bhardwaj2021_m81}, the hosts of repeating FRBs lie firmly in the star-forming/spiral galaxy region. Three non-repeating events (FRBs 20221012A, 20221101B, 20221106A; \citealt{Shannon2024,Sharma2024}), lie within the nominal demarcation for elliptical galaxies, but we note that there is significant overlap between galaxy populations in color-color space (e.g., \citealt{Stern2012,Hainline2014}) and that detailed morphological modeling is required to confirm their elliptical classification.

We thus find that the host of \frb\ is an elliptical galaxy based on robust morphological modeling, marking it as the first of its kind among the population of FRB hosts. The identification of an elliptical host within a transient class that is otherwise dominated by star-forming/disk galaxies bears similarity to transients whose progenitors exhibit a wide range of delay times, leading to their occurrence in both late- and early-type galaxies. This includes ULXs, a third of which are found in early-type galaxies (e.g., \citealt{Plotkin2014}), and short GRBs, where it is estimated that $20-40\%$ occur in elliptical galaxies \citep{Fong2013}. 

Given the largely star-forming host population of FRBs, we briefly consider whether the morphological classification supports that the progenitor of \frb\ formed via a CCSN. Fewer than $1\%$ of CCSNe reside in elliptical galaxies \citep{Irani2022}; such events have been interpreted as ``late'' (200 Myr) explosions following the merger of two intermediate mass ($4-8 \ \rm M_{\odot}$) stars \citep{Zapartas2017} or Ca-rich SNe produced via interacting white dwarfs \citep{Perets2010}, and hence a population distinct from ordinary CCSNe. With nearly 100 FRB hosts discovered to date, the detection of events in rarer environments such as elliptical galaxies becomes more tenable. However, the large projected offset of $\sim 40$ kpc \citep{Shah_R155} would require a hypervelocity ejection with $v \sim 4000 \ \rm km \ s^{-1}$ within a $10$ Myr lifespan; this can be treated as a lower limit since this assumes a purely radial trajectory, and only accounts for projected distance. Consequently, even when considering the small fraction of CCSNe observed in elliptical galaxies, the characteristics of \frb\ and its host strongly support a delayed channel.

\section{Discussion and Conclusions}
\label{sec:conclusions}

We have presented optical through mid-IR observations of the host galaxy of \frb\ at $z=0.1384$, the first repeating FRB localized to a quiescent host, and the first confirmed elliptical galaxy among the population of FRB hosts. Compared to the stellar population properties of FRB hosts, \frb\ is the oldest ($t_m = 11$ Gyr) and most massive ($2 \times 10^{11} \ \rm M_{\odot}$), and shows no evidence for ongoing star formation (log(sSFR) $< -11.8 \ \rm yr^{-1}$), in stark contrast to the bulk of FRB host galaxies that are predominantly found in star-forming environments. Its mass-assembly history further indicates that the majority of mass was formed in the first $\sim$~Gyr, implying long delay times for stellar mass progenitors. 

A comparison to the hosts of other transient classes indicates that \frb's host most closely resembles a subset of hosts for short GRBs, Type Ia supernovae (a proxy for WD AIC), and ULXs. Such channels have previously been proposed as putative progenitors for a small fraction of FRBs based on their localizations to quiescent galaxies which implicate large average delay times between star formation and the transient event \citep{Margalit2019,Li2020,Sridhar2021}.

The discovery of the BNS merger GW170817, along with the overall host demographics of short GRBs, strongly supports a link between short GRBs and BNS mergers. The fraction of BNS mergers that result in stable magnetar remnants remains highly uncertain, however, as it is sensitive to both the unknown equation of state for high-density matter and the mass distribution of BNS systems \citep{Metzger2008,Piro2017}. For a mass distribution that follows that of the Galactic BNS population (e.g., \citealt{Kim2015}), at most a few percent of BNS mergers may leave behind a stable neutron star \citep{Margalit2019b}, which can account for a small fraction of the observed FRB population depending on the FRB active lifetime \citep{Nicholl2017,Margalit2019}. The quiescent host environment of \frb\ therefore lends support for a magnetar progenitor formed through a BNS merger, although as discussed in \citet{Shah_R155}, only $\sim 15\%$ of short GRBs have host-normalized offsets comparable to \frb.

Despite no direct detections of WD AIC, there exists indirect observational evidence for this mechanism, including the presence of young pulsars in globular clusters \citep{Boyle2011}, the low space velocities of recycled millisecond pulsars \citep{Tauris2013}, and the fact that the rate of formation of binary millisecond pulsars can be reconciled with AIC \citep{Hurley2010}. As with BNS mergers, rate estimates for the fraction of AIC leaving behind stable magnetars are highly uncertan \citep{Fryer1999,Kwiatkowski2015}. However, as discussed by \citet{Margalit2019}, highly magnetized neutron stars may be formed via AIC at a rate comparable to that of the BNS magnetar channel either through flux freezing of the initial WD's magnetic field or from a rapidly rotating WD progenitor. Moreover, a typical production timescale of $\sim 10$ Gyr for AIC can account for the occurrence of young neutron stars in GCs \citep{Tauris2013}. We thus find that \frb's quiescent host environment, coupled with a large spatial offset indicative of a possible GC origin \citep{Shah_R155}, lends support for WD AIC as a potential formation channel for \frb. We note that while the transient optical emission from AIC, lasting only a few weeks to months, is expected to be several orders of magnitude fainter than for Type Ia supernovae (e.g., \citealt{LongoMicchi2023}), long-lasting ($\sim$ decades) persistent radio emission from the ejecta interacting with the ambient medium may be more readily detected as multiwavelength counterparts to FRBs \citep{Margalit2019}.

On the other hand, given the frequency ($\sim 30\%$) of ULXs in elliptical galaxies (e.g., \citealt{Plotkin2014}), we suggest an X-ray binary as a possible progenitor for \frb. ULXs have previously been proposed as putative progenitors of FRBs \citep{Sridhar2021}. While ULXs in late-type galaxies are canonically associated with intermediate- or high-mass donor stars, the occurrence of a specific subset of events in elliptical galaxies implicate older, more evolved stellar donors. In this scenario, the expected X-ray luminosity will trace the faint end of the ULX luminosity distribution ($L_X \sim 10^{39} \ \rm erg \ s^{-1}$). With the exception of FRB\,20200120E in the M81 GC ($d \sim 3.6$ Mpc)  \citep{Pearlman2023}, existing X-ray limits for FRBs --- the bulk of which are at $z\gtrsim 0.1$ --- are not sufficiently sensitive to probe such X-ray counterparts \citep{Eftekhari2023,Cook2024}. Indeed, the redshift of \frb\ likely precludes a detection with present-day X-ray facilities. Nevertheless, \frb's elliptical host galaxy, along with a possible GC environment \citep{Shah_R155}, may point to a ULX or X-ray binary progenitor. We note that a ULX origin would distinguish \frb\ from FRB\,20200120E, as X-ray measurements rule out such a scenario for the latter \citep{Pearlman2023}. Conversely, low-mass X-ray binaries remain viable for both sources.

Since the first host associations, investigations into FRB host demographics have offered valuable insights into the origins of FRBs and their possible progenitor systems. Such studies remain in their infancy, however. With the development of interferometric capabilities for various FRB experiments and the promise of hundreds of precisely localized events, the discovery landscape for new and unforeseen hosts and environments presents considerable potential. Indeed, the connection of a few FRBs with remarkable environments, including dwarf galaxies \citep{Chatterjee2017,Niu2022}, a globular cluster \citep{Kirsten2022}, and the elliptical host of \frb, implicate exotic formation channels as well as older stellar populations for some FRBs, and demonstrate that novel environments offer significant constraining power for FRB progenitors. A larger sample of host associations will further uncover intriguing diversity in host environments and may identify interesting subpopulations or correlations with FRB repetition, energetics, or other burst characteristics, contributing to a clearer understanding of FRB origins.

\begin{acknowledgments}

The authors thank Joel Leja and Peter Blanchard for valuable discussions regarding the implementation of \texttt{Prospector}. We thank Steve Schulze and Nabeel Rehemtulla for helping to conduct our Keck observations, and we are grateful to Jennifer Andrews and the Gemini Observatory staff for executing our Gemini observations. T.E. is supported by NASA through the NASA Hubble Fellowship grant HST-HF2-51504.001-A awarded by the Space Telescope Science Institute, which is operated by the Association of Universities for Research in Astronomy, Inc., for NASA, under contract NAS5-26555. Y.D. is supported by the National Science Foundation Graduate Research Fellowship under grant No. DGE-2234667. W.F. gratefully acknowledges support by the National Science Foundation under grant no. AST-2206494 and CAREER grant No. AST-2047919, the David and Lucile Packard Foundation, the Alfred P. Sloan Foundation, and the Research Corporation for Science Advancement through Cottrell Scholar Award \#28284. V.S. is supported by a Fonds de Recherche du Quebec - Nature et Technologies~(FRQNT) Doctoral Research Award. S.S. is a Northwestern University and University of Chicago Brinson Postdoctoral Fellow. The Fast and Fortunate for FRB Follow-up team acknowledges support from NSF grants AST-1911140, AST-1910471, and AST-2206490. The Fong Group at Northwestern acknowledges support by the National Science Foundation under grant Nos. AST-1909358, AST-2206494, AST-2308182, and CAREER grant No. AST-2047919.

B.C.A. is supported by a FRQNT Doctoral Research Award. M.B is a McWilliams fellow and an International Astronomical Union Gruber fellow. M.B. also receives support from the McWilliams seed grant. E.F. and S.S.P. are supported by the National Science Foundation (NSF) grant AST-2407399. J.W.T.H. and the AstroFlash research group acknowledge support from a Canada Excellence Research Chair in Transient Astrophysics (CERC-2022-00009); the European Research Council (ERC) under the European Union’s Horizon 2020 research and innovation programme (`EuroFlash'; Grant agreement No. 101098079); and an NWO-Vici grant (`AstroFlash'; VI.C.192.045). V.M.K. holds the Lorne Trottier Chair in Astrophysics \& Cosmology, a Distinguished James McGill Professorship, and receives support from an NSERC Discovery grant (RGPIN 228738-13). C.L. is supported by NASA through the NASA Hubble Fellowship grant HST-HF2-51536.001-A awarded by the Space Telescope Science Institute, which is operated by the Association of Universities for Research in Astronomy, Inc., under NASA contract NAS5-26555. K.W.M. holds the Adam J. Burgasser Chair in Astrophysics. K.N. is an MIT Kavli Fellow. A.P. is funded by the NSERC Canada Graduate Scholarships - Doctoral program. A.B.P. is a Banting Fellow, a McGill Space Institute~(MSI) Fellow, and a FRQNT postdoctoral fellow. Z.P. is supported by an NWO Veni fellowship (VI.Veni.222.295). M.W.S. acknowledges support from the Trottier Space Institute Fellowship program. P.S. acknowledges the support of an NSERC Discovery Grant (RGPIN-2024-06266). K.S. is supported by the NSF Graduate Research Fellowship Program. FRB research at UBC is supported by an NSERC Discovery Grant and by the Canadian Institute for Advanced Research.  The baseband recording system on CHIME/FRB is funded in part by a CFI John R. Evans Leaders Fund award to IHS.

We acknowledge that CHIME and the \kkoname{} Outrigger (KKO) are built on the traditional, ancestral, and unceded territory of the Syilx Okanagan people. \kkoname{} is situated on land leased from the Imperial Metals Corporation. We are grateful to the staff of the Dominion Radio Astrophysical Observatory, which is operated by the National Research Council of Canada. CHIME operations are funded by a grant from the NSERC Alliance Program and by support from McGill University, University of British Columbia, and University of Toronto. CHIME/FRB Outriggers are funded by a grant from the Gordon \& Betty Moore Foundation. We are grateful to Robert Kirshner for early support and encouragement of the CHIME/FRB Outriggers Project, and to Dusan Pejakovic of the Moore Foundation for continued support. CHIME was funded by a grant from the Canada Foundation for Innovation (CFI) 2012 Leading Edge Fund (Project 31170) and by contributions from the provinces of British Columbia, Québec and Ontario. The CHIME/FRB Project was funded by a grant from the CFI 2015 Innovation Fund (Project 33213) and by contributions from the provinces of British Columbia and Québec, and by the Dunlap Institute for Astronomy and Astrophysics at the University of Toronto. Additional support was provided by the Canadian Institute for Advanced Research (CIFAR), the Trottier Space Institute at McGill University, and the University of British Columbia. The CHIME/FRB baseband recording system is funded in part by a CFI John R. Evans Leaders Fund award to IHS.

Based on observations obtained at the international Gemini Observatory (Program ID: GN-2024A-LP-110), a program of NOIRLab, which is managed by the Association of Universities for Research in Astronomy (AURA) under a cooperative agreement with the National Science Foundation on behalf of the Gemini Observatory partnership: the National Science Foundation (United States), National Research Council (Canada), Agencia Nacional de Investigaci\'{o}n y Desarrollo (Chile), Ministerio de Ciencia, Tecnolog\'{i}a e Innovaci\'{o}n (Argentina), Minist\'{e}rio da Ci\^{e}ncia, Tecnologia, Inova\c{c}\~{o}es e Comunica\c{c}\~{o}es (Brazil), and Korea Astronomy and Space Science Institute (Republic of Korea).

This research was supported in part through the computational resources and staff contributions provided for the Quest high performance computing facility at Northwestern University, which is jointly supported by the Office of the Provost, the Office for Research, and Northwestern University Information Technology. Some of the data presented herein were obtained at the W. M. Keck Observatory, which is operated as a scientific partnership among the California Institute of Technology, the University of California, and the National Aeronautics and Space Administration. The Observatory was made possible by the generous financial support of the W. M. Keck Foundation. The authors wish to recognize and acknowledge the very significant cultural role and reverence that the summit of Maunakea has always had within the indigenous Hawaiian community. We are most fortunate to have the opportunity to conduct observations from this mountain. W. M. Keck Observatory access was supported by Northwestern University and the Center for Interdisciplinary Exploration and Research in Astrophysics (CIERA).

\end{acknowledgments}

\facilities{Gemini
(GMOS), Keck I (LRIS), Las Cumbres Observatory 
}

\software{
\texttt{Astropy} \citep{astropy:2013, astropy:2018, astropy:2022},
\texttt{Dynesty} \citep{Speagle2020},
\texttt{FFFF-PZ} \citep{Coulter2022, Coulter2023},
\texttt{GALFIT} \citep{Peng2002}, 
\texttt{imexam} \citep{imexam}
\texttt{matplotlib} \citep{matplotlib},
\texttt{numpy} \citep{numpy},
\texttt{photutils} \citep{photutils},
\texttt{Prospector} \citep{Johnson2017,Johnson2021},
\texttt{PypeIt} \citep{Prochaska2020},
\texttt{Python-fsps} \citep{Conroy2009,Conroy2010}, 
\texttt{SAOImageDS9} \citep{DS9},
\texttt{scipy} \citep{scipy}
}

\appendix
\section{Stellar Population Modeling with Prospector}
\label{sec:appendix_prospector}

 We use the Flexible Stellar Population Synthesis (FSPS; \citealt{Conroy2009,Conroy2010}) library accessed through the \texttt{PYTHON-FSPS} interface \citep{FSPSpython} to build the stellar population models and the \texttt{dynesty} dynamic nested sampling package \citep{Speagle2020} to jointly fit the host photometry and spectroscopy. We utilize the Gemini/GMOS spectrum given its wide wavelength coverage and high signal-to-noise, and correct both the spectrum and photometry for Milky Way extinction and include the $1\sigma$ photometric uncertainties and error spectrum.

For our fits, we adopt a nonparametric star formation history (SFH) with 8 temporal bins and a continuity prior which utilizes a Student-$t$ prior on the log of the SFR ratio in adjacent bins, as in previous analyses of FRB hosts \citep{Gordon2023}. We additionally employ a Kroupa initial mass function (IMF; \citealt{Kroupa2001}) and the dust attenuation law from \citealt{Kriek13}. To ensure we are sampling from realistic mass and metallicity priors, we assume a Gaussian scatter around the mass-metallicity relation of \citealt{Gallazzi2005} with a standard deviation equal to twice the measured scatter. Finally, we include a pixel outlier model to marginalize over residual sky lines and a two-component mid-IR AGN model as described in \citealt{Leja2018}.

\section{\texttt{GALFIT} Modelling}
\label{sec:appendix_galfit}
To perform the \texttt{GALFIT} modeling in Section~\ref{sec:galfit}, we first identify stars in the image using the \texttt{imexam} package \citep{imexam} and generate the point-spread function (PSF) with \texttt{photutils} \citep{photutils}. From the GMOS image, we create a $1\arcmin\times 1\arcmin$ cutout centered around the host galaxy and produce a segmentation map using \texttt{photutils} to mask other sources in the field. The cutout, PSF profile, and the mask are fed as input to \texttt{GALFIT} along with our initial guesses for the best-fit parameters of a single S\'ersic profile.

\bibliography{references}{}
\bibliographystyle{aasjournal}

\end{document}

%% file: affiliations.tex
\newcommand{\CIERA}{\affiliation{Center for Interdisciplinary Research in Astronomy, Northwestern University, 1800 Sherman Avenue, Evanston, IL 60201, USA }}

\newcommand{\NU}{\affiliation{Department of Physics and Astronomy, Northwestern University, Evanston, IL 60208, USA}}

\newcommand{\Uch}
{\affiliation{Department of Astronomy and Astrophysics, University of Chicago, William Eckhart Research Center, 5640 South Ellis Avenue, Chicago, IL 60637, USA}}

\newcommand{\UCSC}{\affiliation{Department of Astronomy and Astrophysics, University of California Santa Cruz, 1156 High Street, Santa Cruz, CA 95064, USA}}

\newcommand{\IPMU}{\affiliation{Kavli Institute for the Physics and Mathematics of the Universe (Kavli IPMU), 5-1-5 Kashiwanoha, Kashiwa, 277-8583, Japan}}

\newcommand{\NAOJ}{\affiliation{Division of Science, National Astronomical Observatory of Japan, 2-21-1 Osawa, Mitaka, Tokyo 181-8588, Japan}}

\newcommand{\MU}{\affiliation{Department of Physics, McGill University, 3600 rue University, Montr\'eal, QC H3A 2T8, Canada}}

\newcommand{\Trottier}{\affiliation{Trottier Space Institute, McGill University, 3550 rue University, Montr\'eal, QC H3A 2A7, Canada}}

\newcommand{\CMU}{\affiliation{McWilliams Center for Cosmology \& Astrophysics, Department of Physics, Carnegie Mellon University, Pittsburgh, PA 15213, USA}}

\newcommand{\UVA}
{\affiliation{Anton Pannekoek Institute for Astronomy, University of Amsterdam, Science Park 904, 1098 XH Amsterdam, The Netherlands}}

\newcommand{\ASTRON}
{\affiliation{ASTRON, Netherlands Institute for Radio Astronomy, Oude Hoogeveensedijk 4, 7991 PD Dwingeloo, The Netherlands
}}

\newcommand{\MITK}
{\affiliation{MIT Kavli Institute for Astrophysics and Space Research, Massachusetts Institute of Technology, 77 Massachusetts Ave, Cambridge, MA 02139, USA}}

\newcommand{\MITP}
{\affiliation{Department of Physics, Massachusetts Institute of Technology, 77 Massachusetts Ave, Cambridge, MA 02139, USA}}

\newcommand{\CCAPS}{\affiliation{Cornell Center for Astrophysics and Planetary Science, Cornell University, Ithaca, NY 14853, USA}}

\newcommand{\DI}
{\affiliation{Dunlap Institute for Astronomy and Astrophysics, 50 St. George Street, University of Toronto, ON M5S 3H4, Canada}}

\newcommand{\DAA}
{\affiliation{David A. Dunlap Department of Astronomy and Astrophysics, 50 St. George Street, University of Toronto, ON M5S 3H4, Canada}}

\newcommand{\STSCI}
{\affiliation{Space Telescope Science Institute, 3700 San Martin Drive, Baltimore, MD 21218, USA}}

\newcommand{\WVUPHAS}
{\affiliation{Department of Physics and Astronomy, West Virginia University, PO Box 6315, Morgantown, WV 26506, USA }}

\newcommand{\WVUGWAC}
{\affiliation{Center for Gravitational Waves and Cosmology, West Virginia University, Chestnut Ridge Research Building, Morgantown, WV 26505, USA}}

\newcommand{\UCB}
{\affiliation{Department of Astronomy, University of California, Berkeley, CA 94720, United States}}

\newcommand{\YORK}
{\affiliation{Department of Physics and Astronomy, York University, 4700 Keele Street, Toronto, ON MJ3 1P3, Canada}}

\newcommand{\PI}
{\affiliation{Perimeter Institute of Theoretical Physics, 31 Caroline Street North, Waterloo, ON N2L 2Y5, Canada}}

\newcommand{\UBC}
{\affiliation{Department of Physics and Astronomy, University of British Columbia, 6224 Agricultural Road, Vancouver, BC V6T 1Z1 Canada}}

\newcommand{\UCHILE}
{\affiliation{Department of Electrical Engineering, Universidad de Chile, Av. Tupper 2007, Santiago 8370451, Chile}}

%% file: authors.tex
\author[0000-0003-0307-9984]{T.~Eftekhari}
\altaffiliation{NHFP Einstein Fellow}
\CIERA

\author[0000-0002-9363-8606]{Y.~Dong \begin{CJK*}{UTF8}{gbsn}(董雨欣)\end{CJK*}}
\CIERA
\NU

\author[0000-0002-7374-935X]{W.~Fong}
\CIERA
\NU

\author[0000-0002-4823-1946]{V.~Shah}
\MU
\Trottier

\author[0000-0003-3801-1496]{S.~Simha}
\CIERA
\Uch


\author[0000-0001-5908-3152]{B.~C.~Andersen}
\MU
\Trottier

\author[0000-0002-3980-815X]{S.~Andrew}
\MITK
\MITP

\author[0000-0002-3615-3514]{M.~Bhardwaj}
\CMU

\author[0000-0003-2047-5276]{T.~Cassanelli}
\UCHILE

\author[0000-0002-2878-1502]{S.~Chatterjee}
\CCAPS

\author[0000-0003-4263-2228]{D.~A.~Coulter}
\STSCI

\author[0000-0001-8384-5049]{E.~Fonseca}
\WVUPHAS
\WVUGWAC

\author[0000-0002-3382-9558]{B.~M.~Gaensler}
\UCSC
\DAA
\DI

\author[0000-0002-5025-4645]{A.~C.~Gordon}
\CIERA
\NU

\author[0000-0003-2317-1446]{J.~W.~T.~Hessels}
\MU
\Trottier
\UVA
\ASTRON

\author[0000-0003-2405-2967]{A.~L.~Ibik}
\DI
\DAA

\author[0000-0003-3457-4670]{R.~C.~Joseph}
\MU
\Trottier

\author[0009-0007-5296-4046]{L.~A.~Kahinga}
\UCSC

\author[0000-0001-9345-0307]{V.~Kaspi}
\MU 
\Trottier

\author[0009-0008-6166-1095]{B.~Kharel}
\WVUPHAS
\WVUGWAC

\author[0000-0002-5740-7747]{C.~D.~Kilpatrick}
\CIERA

\author[0000-0003-2116-3573]{A.~E.~Lanman}
\MITK
\MITP

\author[0000-0002-5857-4264]{M.~Lazda}
\DAA
\DI

\author[0000-0002-4209-7408]{C. Leung}
\altaffiliation{NHFP Einstein Fellow}
\UCB

\author[0000-0002-7866-4531]{C.~Liu}
\CIERA
\NU

\author[0000-0003-4584-8841]{L.~Mas-Ribas}
\UCSC

\author[0000-0002-4279-6946]{K.~W.~Masui}
\MITK
\MITP

\author[0000-0001-7348-6900]{R.~Mckinven}
\MU
\Trottier

\author[0000-0002-0772-9326]{J.~Mena-Parra}
\DAA
\DI

\author[0000-0001-9515-478X]{A.~A.~Miller}
\CIERA
\NU

\author[0000-0003-0510-0740]{K.~Nimmo}
\MITK

\author[0000-0002-8897-1973]{A.~Pandhi}
\DAA
\DI

\author[0000-0002-8912-0732]{A.~B.~Pearlman}
\altaffiliation{Banting Fellow, McGill Space Institute~(MSI) Fellow, \\ and FRQNT Postdoctoral Fellow.}
\MU
\Trottier

\author[0000-0002-4795-697X]{Z.~Pleunis}
\UVA
\ASTRON

\author[0000-0002-7738-6875]{J.~X.~Prochaska}
\UCSC
\IPMU
\NAOJ

\author[0000-0001-7694-6650]{M.~Rafiei-Ravandi}
\MU

\author[0000-0002-4623-5329]{M.~Sammons}
\MU
\Trottier

\author[0000-0002-7374-7119]{P.~Scholz}
\YORK
\DI 

\author[0000-0002-6823-2073]{K.~Shin}
\MITK
\MITP

\author[0000-0002-2088-3125]{K.~Smith}
\PI

\author[0000-0001-9784-8670]{I.~Stairs}
\UBC

\author[0009-0008-7264-1778]{P.~Swarali Shivraj}
\WVUPHAS
\WVUGWAC